\documentclass[conference,transmag,10pt]{IEEEtran} 
\usepackage[english]{babel}
\usepackage[utf8]{inputenc}
\usepackage[T1]{fontenc}
\usepackage{siunitx}
\usepackage{graphicx}
\graphicspath{{./figs/}}
\usepackage{subcaption}
\usepackage{bbm}
\usepackage{pgfplots}
\pgfplotsset{compat=newest}
\usepgfplotslibrary{external}
\usepackage{tikz}
\usepackage{circuitikz}
\usetikzlibrary{shapes}
\usepackage{algorithm}
\usepackage{algpseudocode}
\usepackage{amsmath}

\usepackage{flushend}  
\usepackage{balance}   
\usepgfplotslibrary{external}
\usepackage{eso-pic}

\catcode`@=11
\def\frownfill{$\scriptscriptstyle\m@th\mathord\frown$}
\def\bow#1{\vbox{\m@th\ialign{##\crcr
      \hfil\frownfill\hfil\crcr\noalign{\kern-0.2\p@\nointerlineskip}
      $\hfil\displaystyle{#1}\hfil$\crcr}}}
\def\bbow#1{\vbox{\m@th\ialign{##\crcr
     \hfil\frownfill\hfil\crcr\noalign{\kern-0.7\p@\nointerlineskip}
     \hfil\frownfill\hfil\crcr\noalign{\kern-0.3\p@\nointerlineskip}
      $\hfil\displaystyle{#1}\hfil$\crcr}}}
\def\widefrownfill{$\m@th\mathord\frown$}
\def\widebow#1{\vbox{\m@th\ialign{##\crcr
      \hfil\widefrownfill\hfil\crcr\noalign{\kern-0.9\p@\nointerlineskip}
      $\hfil\displaystyle{#1}\hfil$\crcr}}}
\def\widebbow#1{\vbox{\m@th\ialign{##\crcr
     \hfil\widefrownfill\hfil\crcr\noalign{\kern-1.8\p@\nointerlineskip}
     \hfil\widefrownfill\hfil\crcr\noalign{\kern-0.9\p@\nointerlineskip}
      $\hfil\displaystyle{#1}\hfil$\crcr}}}

\AddToShipoutPicture*{\footnotesize\sffamily\raisebox{1cm}{\hspace{1.65cm}\fbox{\parbox{\textwidth}{\copyright~2016 IEEE. Personal use of this material is permitted. Permission from IEEE must be obtained for all other uses, in any current or future media, including reprinting/republishing this material for advertising or promotional purposes, creating new collective works, for resale or redistribution to servers or lists, or reuse of any copyrighted component of this work in other works.}}}}\newcommand{\figref}[1]{Fig.~\ref{#1}}

\title{\bfseries\Large Automatic Generation of Equivalent Electrothermal SPICE Netlists from 3D Electrothermal Field Models}
\author{\IEEEauthorblockN{Thorben~Casper\IEEEauthorrefmark{1,2},Herbert~De~Gersem\IEEEauthorrefmark{2},
		Sebastian~Schöps\IEEEauthorrefmark{1,2}}
\IEEEauthorblockA{\IEEEauthorrefmark{1}Graduate School of Computational Engineering,
Technische Universität Darmstadt, 64293 Darmstadt, Germany}
\IEEEauthorblockA{\IEEEauthorrefmark{2}Institut für Theorie Elektromagnetischer Felder, Technische Universität Darmstadt, 64289 Darmstadt, Germany\\
Email: casper@gsc.tu-darmstadt.de, Phone: +4961511624392}}

\begin{document}
\maketitle
\begin{abstract}
Starting from a 3D electrothermal field problem discretized by the Finite Integration Technique, the equivalence to a circuit description is shown by exploiting the analogy to the Modified Nodal Analysis approach. Using this analogy, an algorithm for the automatic generation of a monolithic SPICE netlist is presented. Joule losses from the electrical circuit are included as heat sources in the thermal circuit. The thermal simulation yields nodal temperatures that influence the electrical conductivity. Apart from the used field discretization, this approach applies no further simplifications. An example 3D chip package is used to validate the algorithm.
\end{abstract}

\section{Introduction}
\label{sec:introduction}

	For many electrothermal applications, complex problems need to be solved to calculate the desired quantities of interest. While analytical methods only serve for very simple problems, real world problems require numerical field simulations. Nowadays, Finite Difference (FD), Finite Element (FE) or Finite Volume (FV) methods are well known and frequently applied. On the other hand, when a fast computation is required, it is often recommended to derive a model that only consists of a few lumped elements which allows to use circuit simulators. However, network models are not always sufficiently accurate or they are too complex to be generated by hand compared to models based on partial differential equations. This paper attempts to fill this gap by deriving a circuit netlist from a field model, thereby avoiding any further approximation.
	
	When the first circuit simulators were introduced, the Simulation Program with Integrated Circuit Emphasis~(SPICE) soon became the standard tool for describing and solving circuits using netlists. At that time, the operating temperature of the simulated devices was set globally and did not change according to the operational load of the circuit. However, especially in power electronic applications that involve strongly temperature dependent materials \cite{Marz_2000aa,Kosel_2010aa}, self-heating and thermal coupling between devices become important and therefore the device's temperature changes dynamically. Thus, the concurrent simulation of electrical and thermal circuits was soon developed as a SPICE extension \cite{Vogelsong_1989aa}. To account for the coupling, an extra temperature node was added to the electrical device models \cite{Mawby_2001aa,Hefner_1994aa}. 

\begin{figure}[b!]
		\vspace{-0.5em}
		\centering
		\resizebox{0.8\columnwidth}{!}{\includegraphics{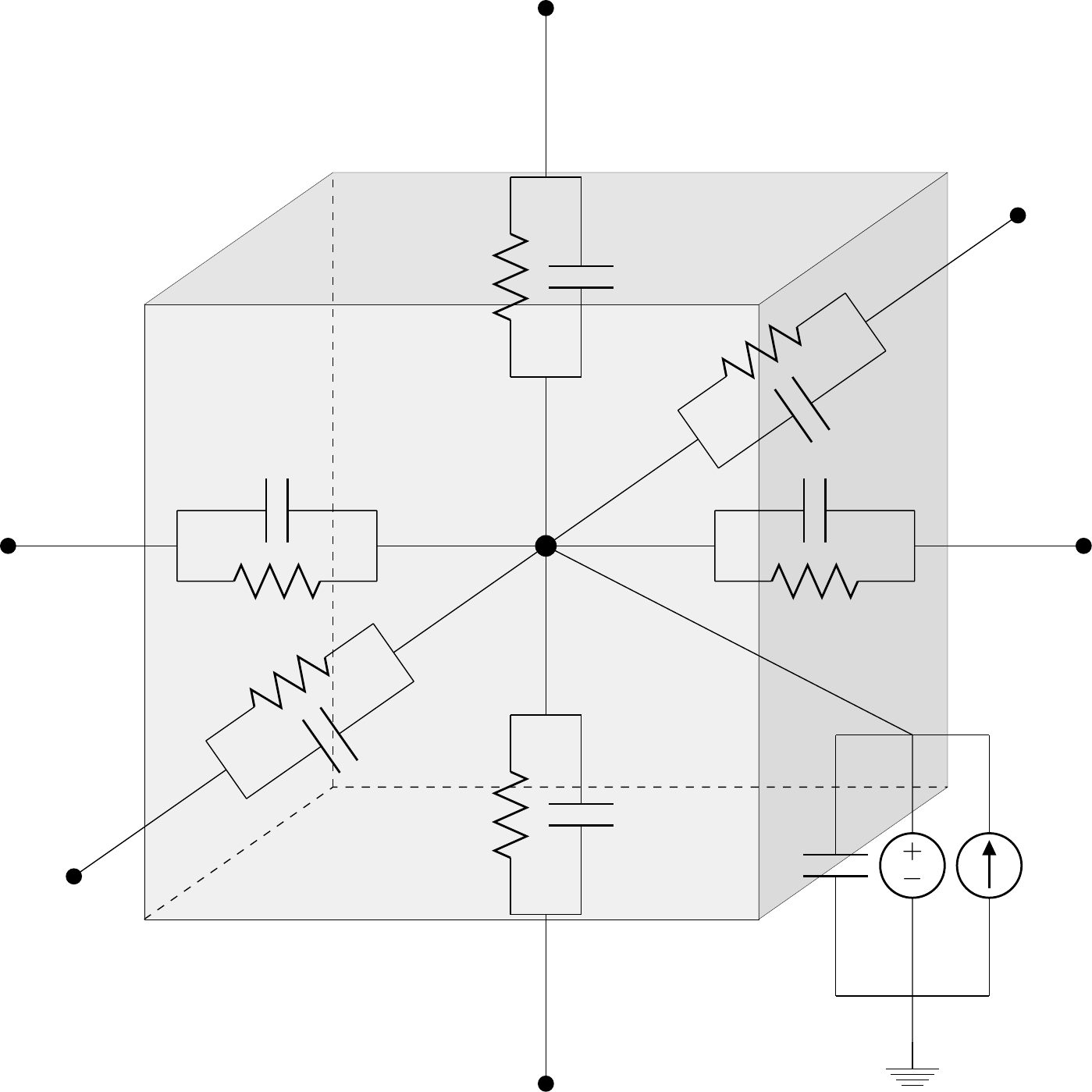}
 }
		\caption{Visualization of the approach to extract lumped elements from a 3D field model.}
		\label{fig:lumpedElementsOnCube}
	\end{figure}

	Traditionally, the circuit topology and the netlist parameters are determined by hand calculations which necessarily neglect nonlinear effects and field inhomogeneities. Nowadays, the thermal behavior is more accurately analyzed by 3D field solvers which allow to couple the resulting temperature distribution iteratively to the electric circuit simulator, known as the relaxation method~\cite{Van-Petegem_1994aa,Chvala_2014aa,Wunsche_1997aa}. On the other hand, the definition of mesh-based equivalent thermal circuits \cite{Simpson_2014ab} motivates the direct method that couples the electrical and thermal circuits without any 3D field solver required. Following the direct approach, this work aims at exploiting the accurate electrothermal field model by automatic and loss-free extraction of an equivalent electrothermal circuit model (as visualized in \figref{fig:lumpedElementsOnCube}). The netlists generated by this approach are therefore an exact representation of the nonlinear FD, FE or FV field discretizations. Moreover, a straightforward coupling with an additional (external) SPICE circuitry is possible and standard techniques of model order reduction for networks~\cite{Hinze_2012ad} can be applied afterwards.
	
	If a 3D electrothermal field model is investigated, occasionally surrounding circuitry is treated by a field-circuit coupled approach \cite{Benderskaya_2004aa,Schops_2013aa}. In some cases, a two-step procedure is executed, i.e., from the field model part, a reduced model is extracted and then included in a circuit model \cite{Wittig_2002aa}. In this paper, a further alternative is proposed, i.e., the full field model is translated into a circuit model. This approach is promising for small device parts with severe electrothermal issues, to be embedded in an overall circuit model.

	The paper is organized as follows: The electrothermal field problem is introduced in section \ref{sec:field_problem}. Then, section~\ref{sec:FIT} presents the discretization by the Finite Integration Technique (FIT). This technique is based on integral-type degrees of freedom and therefore allows a straightforward circuit interpretation as demonstrated in section \ref{sec:circuit_theory}. The details of the automatic SPICE netlist generation are given in section \ref{sec:methodology}. Finally, numeric validation examples are discussed in section \ref{sec:numerical_example} before section \ref{sec:conclusion} concludes the paper. \section{Electrothermal Field Problem}
\label{sec:field_problem}

 We consider an electrothermal problem that couples the electroquasistatic approximation of Maxwell's equations~\cite{Clemens_2004ab} with the transient nonlinear heat equation
	\begin{align}
	-\nabla\cdot\left(\varepsilon\nabla\dot{\varphi}(t)\right)-\nabla\cdot\left(\sigma(T)\nabla\varphi(t)\right) &= 0,
	\label{eq:EQS_continuous}
	\\
	\rho c\dot{T} - \nabla\cdot\left(\lambda(T)\nabla T(t)\right) &= Q_\text{el}(\varphi),
	\label{eq:thermal_continuous}
	\end{align}
	subject to adequate initial and boundary conditions. The given quantities are the Joule heating term ${Q_\text{el}(\varphi) = \sigma(T)\nabla\varphi\cdot\nabla\varphi}$, the time $t$, the temperature $T$ and the electric scalar potential $\varphi$. The material parameters $\varepsilon$, $\sigma$, $\lambda$, $\rho$ and $c$ are the electrical permittivity and conductivity, the thermal conductivity, the volumetric mass density and the specific heat capacity, respectively. All involved materials may be nonlinear, inhomogeneous or anisotropic. Note that the spatial dependencies have been suppressed here and that the temperature dependency of $\rho$ and $c$ is neglected.

\section{Discretization by the Finite Integration Technique}
\label{sec:FIT}

\newcommand{\ve}{\protect\bow{\rm\bf e}}
\newcommand{\vt}{\protect\bow{\rm\bf t}}
\newcommand{\fj}{\protect\bbow{\rm\bf j}}
\newcommand{\fq}{\protect\bbow{\rm\bf q}}
\newcommand{\mbf}{\mathbf}

\newcommand{\G}{\mathbf{G}}
\newcommand{\Ss}{\widetilde{\mathbf{S}}^{\top}}

\begin{figure}[b]
	\centering
	\resizebox{0.5\columnwidth}{!}{	\includegraphics{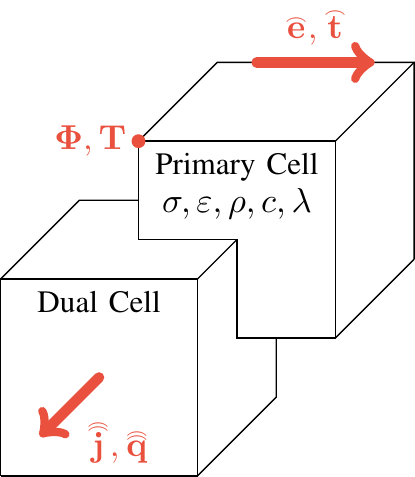}
				
 }
	\caption{Allocation of electrical and thermal quantities at the primary and dual grid.}
	\label{fig:allocation}
\end{figure}

The coupled electrothermal problem given by~\eqref{eq:EQS_continuous}~and~\eqref{eq:thermal_continuous} is discretized in space by the Finite Integration Technique~(FIT) on a staggered 3D hexahedral grid with~$n$~canonically indexed nodes~\cite{Weiland_1977aa, Weiland_1996aa, Clemens_2001ac}. The discretization turns the material relations and the differential operators into material matrices and topological matrices, respectively. The discrete unknowns, i.e., the electric potentials $\mathbf{\Phi}\in\mathbbm{R}^{n}$ and the temperatures $\mathbf{T}\in\mathbbm{R}^{n}$ are associated with the nodes of the primary grid (see \figref{fig:allocation}). The voltage and temperature drops are allocated at the primary edges and resemble differences, i.e., $\ve = -\G\mathbf{\Phi}$ and $\vt = -\G\mbf{T}$, where $\G\in\{-1,0,1\}^{3n\times n}$ is the discrete gradient matrix according to the topology of the primary grid. Electrical currents $\fj$ and heat fluxes~$\fq$~are assigned to the facets of the dual grid. They are accumulated on the dual cells by $\widetilde{\mathbf{S}}\fj$ and $\widetilde{\mathbf{S}}\fq$ where $\widetilde{\mathbf{S}}\in\{-1,0,1\}^{n\times 3n}$ is the discrete divergence matrix determined by the topology of the dual grid. The duality of the staggered grid gives raise to the property $\G=-\Ss$.

Note that the signs of the different entries of ${\G=\begin{bmatrix}\mathbf{P}_x & \mathbf{P}_y & \mathbf{P}_z\end{bmatrix}\hspace{-1pt}\vspace{-1pt}^{\top}}$ (and thus also~$\widetilde{\mathbf{S}}$) define the orientation of the grid's edges. Here, their direction is chosen to be the same as the direction of the coordinate axes. The $n\times n$ sub-matrices $\mathbf{P}_\xi$ with $\xi\in\left\{x,y,z\right\}$ are therefore the discrete representation of the spatial derivatives $\frac{\mathrm{d}}{\mathrm{d}x}$, $\frac{\mathrm{d}}{\mathrm{d}y}$ and $\frac{\mathrm{d}}{\mathrm{d}z}$. This results in $-1$ entries on the main diagonals of $\mathbf{P}_\xi$ and in $+1$ entries on one of the super-diagonals. The discrete gradient matrix~$\G$ resembles an incidence matrix as used in circuit theory that is more structured than a typical circuit incidence matrix. Both matrices share the property that their column sums are zero.

The discrete unknowns at the primary grid are related to the ones at the dual grid by material matrices, i.e., the electrical conductance matrix $\mathbf{M}_{\sigma}\in\mathbbm{R}^{3n\times 3n}$, the electrical capacitance matrix $\mathbf{M}_{\varepsilon}\in\mathbbm{R}^{3n\times 3n}$, the thermal conductance matrix $\mathbf{M}_{\lambda}\in\mathbbm{R}^{3n\times 3n}$ and the thermal capacitance matrix $\mathbf{M}_{\rho c}\in\mathbbm{R}^{3n\times 3n}$. The electrical currents, displacement currents, heat fluxes and stored heats are then given by
\begin{equation*}
	\fj=\mathbf{M}_{\sigma}\ve, \,\,\, \protect\bbow{\rm\bf d}=\mathbf{M}_{\varepsilon}\ve, \,\,\, \fq=\mathbf{M}_{\lambda}\vt \,\,\, \text{and} \,\,\, \mathbf{Q}=\mathbf{M}_{\rho c}\dot{\mathbf{T}},
\end{equation*}
respectively. In case of a mutually orthogonal grid pair, the material matrices are diagonal. Then, each primary edge crosses the corresponding dual facet perpendicularly. The primary edge and facet as well as the primary nodes and dual cells are indexed pairwise identically. This gives the entries of $\mathbf{M}_{\sigma}$, $\mathbf{M}_{\varepsilon}$, $\mathbf{M}_{\lambda}$ and $\mathbf{M}_{\rho c}$ as 
\begin{alignat*}{2}
	&\mbf{M}_{\sigma;j,j} = \frac{\overline{\sigma}_j|\widetilde{A}_j|}{\left|L_j\right|},
	\quad
	&&\mbf{M}_{\varepsilon;j,j} = \frac{\overline{\varepsilon}_j|\widetilde{A}_j|}{\left|L_j\right|},\\
	&\mbf{M}_{\lambda;j,j} = \frac{\overline{\lambda}_j|\widetilde{A}_j|}{\left|L_j\right|}
	\quad
	\text{and}
	\quad
	&&\mbf{M}_{\rho c;i,i} ={\overline{\rho c}_i|\widetilde{V}_i|},
\end{alignat*}
where $\left|L_j\right|$ is the length of the primary edge $L_j$, $|\widetilde{A}_j|$ is the area of the dual facet $\widetilde{A}_j$ and $|\widetilde{V}_i|$ is the volume of dual cell $\widetilde{V}_i$. The material parameters $\overline{\sigma}_j$, $\overline{\varepsilon}_j$, $\overline{\lambda}_j$ and $\overline{\rho c}_i$ are found by averaging the corresponding parameters, cf. \cite{Clemens_2001aa,Casper_2016aa}. Note that the counter $i$ addresses all primary nodes (or dual volumes) while $j$ addresses all primary edges (or dual facets).

The applied averaging scheme depends on the allocation of the materials on the grid. In this paper, the material properties are allocated at the primary grid cells, i.e., each primary cell contains a homogeneous material. This gives rise to the so-called staircase approximation. Although this paper adopts this approximation for reasons of conciseness, partially filled cells \cite{Schauer_2003aa} and conforming techniques \cite{Clemens_2002ad} are preferred. According to the chosen allocation scheme (see \figref{fig:allocation}), the averaging of a conductance for a primary-edge, dual-facet pair involves four surrounding primary volumes resulting in the average of the four involved conductivities. For an equidistant grid, this gives
\begin{equation}
	\overline{\sigma}_j = \frac{1}{4}\sum_{p=1}^{4}\sigma_p,
	\label{eq:average_sigma}
\end{equation}
with $p$ being a local counter over the four primary volumes $V_p$ that share the edge $L_j$. Note that this averaging scheme can easily be generalized for non-equidistant grids by taking the different cell sizes into account.

\begin{figure}[b]
	\vspace{-0.5em}
	\centering
	\resizebox{\columnwidth}{!}{	\includegraphics{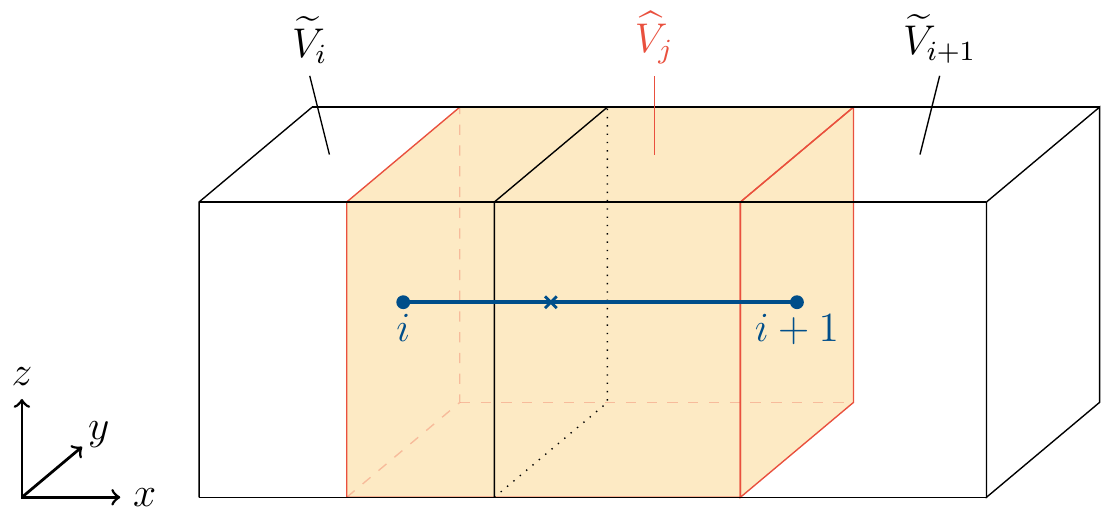}
	 }
	\vspace{-1.5em}
	\caption{Visualization of the shifted volume $\widehat{V}_j$.}
	\label{fig:shiftedVolume}
\end{figure}

The heat powers generated by the thermal losses from the electroquasistatic problem are calculated by
\begin{equation*}
	\widehat{\mathbf{Q}}_\text{el} = \protect\bow{\rm\bf e}\odot\protect\bbow{\rm\bf j},
\end{equation*}
with the component-wise Hadamard product $\odot$ and the vector $\widehat{\mathbf{Q}}_\text{el}\in\mathbbm{R}^{3n\times 1}$ that allocates the thermal losses at the \emph{shifted cells} $\widehat{V}_j$ with the volume ${|\widehat{V}_j| = |\widetilde{A}_j||L_j|}$ as shown in~\figref{fig:shiftedVolume}. The shifted cells consist of two half dual cells sharing a common dual facet and are indexed according to the dual facets.

To include all calculated heat contributions as sources in the heat equation, they need to be integrated from the shifted volumes $\widehat{V}_j$ to the neighboring dual cells $\widetilde{V}_i$. This relation is given by
\begin{equation*}
	Q_{\text{el},i} = \sum_{p=1}^6 \frac{|\widetilde{V}_i|}{2|\widehat{V}_p|}\widehat{Q}_{\text{el},p},
\end{equation*}
where $p$ is a local counter that loops over all shifted cells that intersect the dual cell $i$. To sum up all contributions of the shifted volumes to the dual cell $\widetilde{V}_i$, the incidence matrix ${\mathbf{P}_\text{Q}\in\{0,1\}^{n\times 3n}}$ is defined. Then, the Joule losses $\mathbf{Q}_\text{el}\in\mathbbm{R}^{n}$ on the dual cells become
\begin{equation}
	\mathbf{Q}_\text{el} = \frac{1}{2}\mathbf{\widetilde{D}}_V\mathbf{P}_\text{Q}\mathbf{\widehat{D}_V}^{-1}\widehat{\mathbf{Q}}_\text{el}.
	\label{eq:Qel_definition}
\end{equation}
Here, $\mathbf{\widetilde{D}}_V\in\mathbbm{R}^{n\times n}$ and $\mathbf{\widehat{D}_V}\in\mathbbm{R}^{3n\times 3n}$ are diagonal matrices containing the dual volumes $|\widetilde{V}_i|$ and the shifted volumes $|\widehat{V}_j|$, respectively.

Finally, the topological operators $\widetilde{\mathbf{S}}$ where $\widetilde{\mathbf{S}}^{\top} = -\G$, the material matrices $\mathbf{M}_{\sigma}$, $\mathbf{M}_{\varepsilon}$, $\mathbf{M}_{\lambda}$ and $\mathbf{M}_{\rho c}$ and the heat powers $\mathbf{Q}_\text{el}$ are utilized to express the discrete counterpart of \eqref{eq:EQS_continuous} and \eqref{eq:thermal_continuous}, i.e.,
\begin{align}
\widetilde{\mathbf{S}}\mathbf{M}_{\varepsilon}\widetilde{\mathbf{S}}^{\top}\dot{\mathbf{\Phi}}+\widetilde{\mathbf{S}}\mathbf{M}_{\sigma}(\mathbf{T})\widetilde{\mathbf{S}}^{\top}\mathbf{\Phi} &= \mathbf{0},
\label{eq:EQS_FIT}
\\
\mathbf{M}_{\rho c} \dot{\mathbf{T}} + \widetilde{\mathbf{S}}\mathbf{M}_{\lambda}(\mathbf{T})\widetilde{\mathbf{S}}^{\top} \mathbf{T} &= \mathbf{Q}_\text{el}(\mathbf{\Phi}). 
\label{eq:thermal_FIT}
\end{align}
The degrees of freedom are the discrete temperature vector $\mathbf{T} = \mathbf{T}(t)$ and the electric potential vector $\mathbf{\Phi}=\mathbf{\Phi}(t)$. For both the electric and thermal part of the system, adequate initial and boundary conditions need to be defined. As stated here, \eqref{eq:EQS_FIT} includes the definition of homogeneous Dirichlet boundary nodes while \eqref{eq:thermal_FIT} is subject to homogeneous Neumann (adiabatic) conditions throughout this paper. Subsequently, the time is discretized by any standard scheme, e.g. the backward Euler method. \section{Circuit Theory}
\label{sec:circuit_theory}
The equivalence between the discretized FIT formulations \eqref{eq:EQS_FIT} and \eqref{eq:thermal_FIT} and the Modified Nodal Analysis~(MNA)~\cite{Ho_1975aa,Gunther_2005aa} becomes obvious when the latter is derived from the Maxwell equations along a similar reasoning as in section \ref{sec:FIT}. First, Kirchhoff's Voltage Law (KVL) and Kirchhoff's Current Law (KCL) are derived directly from Faraday's law and the current continuity equation, respectively. Then, they are assembled together with the branch relations to give a circuit formulation. From this, the electric and thermal equivalences to the FIT formulation are obtained.

\subsection{Kirchhoff's Voltage Law (KVL)}
In the electroquasistatic case, Faraday's law reads
\begin{equation*}
	\oint_{\partial A_\text{loop}} \vec{E}\cdot\text{d}\vec{s} = 0,
\end{equation*}
where $A_\text{loop}$ is the surface enclosed by the loop such that $\partial A_\text{loop} = \cup_{j=1}^{b} L_j$, with $b$ branches $L_j$. Loops in a circuit closely resemble loops in the primary grids, i.e., a loop is a collection of edges and nodes forming a closed path. The integral form of Faraday's law directly breaks down in an addition of voltage drops $V_j$ along the branches $L_j$ giving
\begin{equation}
	\oint_{\partial A_\text{loop}} \vec{E}\cdot\text{d}\vec{s} = \sum_{j=1}^{b}\int_{L_j}\vec{E}\cdot\text{d}\vec{s} = \sum_{j=1}^{b}V_j = 0.
	\label{eq:Faraday_loop}
\end{equation}
In MNA, \eqref{eq:Faraday_loop} is implicitly fulfilled by defining the $n$ nodal voltages $v_i$ and expressing the branch voltages $V_j$ by 
\begin{equation*}
	\mathbf{V} = \mathbf{A}\hspace{-2pt}^{\top}\mathbf{v},
\end{equation*}
where $\mathbf{A}\in\{-1,0,1\}^{n\times b}$ is the circuit incidence matrix. In analogy to the FIT matrix $\G$ (cf. section \ref{sec:FIT}), the entries $a_{ij}$ of $\mathbf{A}$ have to be subject to a convention. Let us define $a_{ij} = +1$ if branch $L_j$ is directed away from node $i$ and $a_{ij} = -1$ if branch $L_j$ is directed towards node $i$. If branch $L_j$ is not directly connected to node $i$, the corresponding entry $a_{ij}$ is zero. The nodal voltages $\mathbf{v}$ are the circuit counterpart of the electric potentials $\mathbf{\Phi}$ in the FIT formulation. Note that in the electrical case, at least one of the $n$ potentials needs to be chosen as the reference potential (ground) to ensure uniqueness of the solution.

\subsection{Kirchhoff's Current Law (KCL)}
The KCL is derived from the current continuity equation
\begin{equation}
	\int_{\partial V}\vec{J}\cdot\text{d}\vec{A} = \int_V \dot{\varrho}\,\text{d}V,
	\label{eq:current_continuity}
\end{equation}
where $\varrho(\ensuremath{ \vec{r} },t)$ is the charge density and $V$ an arbitrary cell. In analogy to FIT, we consider a cell $\widetilde{V}_i$ around a node $i$ of the circuit. Furthermore, we assume that the total charge in $\widetilde{V}_i$ is zero. This means that capacitive charges are located on branches that are either fully outside or fully inside $\widetilde{V}_i$. Then, \eqref{eq:current_continuity} becomes
\begin{equation*}
	\int_{\partial \widetilde{V}_i} \vec{J}\cdot\text{d}\vec{A} = 0,
\end{equation*}
with the boundary~$\partial \widetilde{V}_i$ of~$\widetilde{V}_i$. Assuming that a finite number $s$ of conductors with cross-sectional areas $\widetilde{A}_j$ and currents $I_j$ are leaving this cell, KCL is obtained as
\begin{equation*}
	\sum_{j=1}^s I_j = \sum_{j=1}^s \int_{\widetilde{A}_j}\vec{J}\cdot\text{d}\vec{A} = 0.
\end{equation*}
Using the definition of the circuit incidence matrix,
\begin{equation*}
	\mathbf{A}\mathbf{I} = \mathbf{0}
\end{equation*}
states KCL for every node in the network. Note that $\mathbf{0}$ denotes a vector of zeros of suitable dimension.

\subsection{Branch Relations}

Due to the nature of the considered problem, we limit ourselves to resistances, capacitances and current sources as branch elements. With $b_\text{R}$ resistive branches, $b_\text{C}$ capacitive branches and $b_\text{I}$ current sources, $\mathbf{A}$ can be divided into different blocks representing these elements \cite{Gunther_2005aa}. We therefore obtain
\begin{equation*}
	\mathbf{A} = \left[\mathbf{A}_\text{R} \quad \mathbf{A}_\text{C} \quad \mathbf{A}_\text{I}\right],
\end{equation*}
where $\mathbf{A}_\text{R}\in\{-1,0,1\}^{n\times b_\text{R}}$, $\mathbf{A}_\text{C}\in\{-1,0,1\}^{n\times b_\text{C}}$ and $\mathbf{A}_\text{I}\in\{-1,0,1\}^{n\times b_\text{I}}$. The vector of currents is partitioned accordingly, i.e.,
\begin{equation*}
	\mathbf{I}^{\top} = \left[\mathbf{I}_\text{R}^{\top} \quad \mathbf{I}_\text{C}^{\top} \quad \mathbf{I}_\text{I}^{\top}\right],
\end{equation*}
with the blocks $\mathbf{I}_\text{R}\in\mathbbm{R}^{b_\text{R}}$, $\mathbf{I}_\text{C}\in\mathbbm{R}^{b_\text{C}}$ and $\mathbf{I}_\text{I}\in\mathbbm{R}^{b_\text{I}}$. Similarly, with $\mathbf{V}_\text{R}\in\mathbbm{R}^{b_\text{R}}$, $\mathbf{V}_\text{C}\in\mathbbm{R}^{b_\text{C}}$ and $\mathbf{V}_\text{I}\in\mathbbm{R}^{b_\text{I}}$, the branch voltage vector is partitioned as
\begin{equation*}
	\mathbf{V}^{\top} = \left[\mathbf{V}_\text{R}^{\top} \quad \mathbf{V}_\text{C}^{\top} \quad \mathbf{V}_\text{I}^{\top}\right].
\end{equation*}
With these definitions, KVL and KCL are expressed by
\begin{eqnarray}
	\mathbf{V}_\text{R} = \mathbf{A}_\text{R}^{\top}\mathbf{\mathbf{v}}, \quad \mathbf{V}_\text{C} = \mathbf{A}_\text{C}^{\top}\mathbf{\mathbf{v}}, \quad \mathbf{V}_\text{I} = \mathbf{A}_\text{I}^{\top}\mathbf{\mathbf{v}}
	\label{eq:voltage_definition}\\	
	\text{and}\quad\mathbf{A}_\text{R}\mathbf{I}_\text{R} + \mathbf{A}_\text{C}\mathbf{I}_\text{C} + \mathbf{A}_\text{I}\mathbf{I}_\text{I} = \mathbf{0}.
	\label{eq:MNA_KCL}
\end{eqnarray}

The introduction of the diagonal conductance matrix $\mathbf{G}\in\mathbbm{R}^{b_\text{R}\times b_\text{R}}$ and the diagonal capacitance matrix ${\mathbf{C}\in\mathbbm{R}^{b_\text{C}\times b_\text{C}}}$ with the corresponding values on the diagonal leads to the branch relations
\begin{equation*}
	\mathbf{I}_\text{R} = \mathbf{G}\mathbf{V}_\text{R}, \quad \mathbf{I}_\text{C} = \mathbf{C}\dot{\mathbf{V}}_\text{C} \quad \text{and} \quad \mathbf{I}_\text{I} = \mathbf{I}_\text{s}(t),
\end{equation*}
where $\mathbf{I}_\text{I} = \mathbf{I}_\text{s}(t)$ is an arbitrary source current.

The combination of KVL, KCL and the branch relations gives the MNA formulation
\begin{equation}
	\label{eq:MNA_result}
	\mathbf{A}_\text{C}\mathbf{C}\mathbf{A}_\text{C}^{\top}\dot{\mathbf{v}} + \mathbf{A}_\text{R}\mathbf{G}\mathbf{A}_\text{R}^{\top}\mathbf{v} = -\mathbf{A}_\text{I}\mathbf{I}_\text{s}(t).
\end{equation}

\subsection{Circuit Formulation and Electrical Equivalences}
For the EQS case, when comparing the structure of~\eqref{eq:EQS_FIT}~and~\eqref{eq:thermal_FIT} with \eqref{eq:MNA_result}, many equivalences are found. Thanks to the equally chosen orientation of the edges in the FIT grid in comparison to the edges in the MNA, the discrete field formulation can be interpreted as a circuit formulation by setting
\begin{equation}
	\begin{aligned}
		\quad \mathbf{A}_\text{C}^\text{el} = \mathbf{A}_\text{R}^\text{el} = \widetilde{\mathbf{S}}, \quad \mathbf{C}^\text{el} = \mathbf{M}_{\varepsilon}, \quad \mathbf{G}^\text{el} &= \mathbf{M}_{\sigma}, \\
		\quad \mathbf{I}_\text{s}^\text{el} = \mathbf{0} \,\,\,\, \text{and} \,\,\,\hspace{0.5pt} \mathbf{\mathbf{v}}^\text{el} &= \mathbf{\Phi},
		\label{eq:EQS_equivalence}
	\end{aligned}
\end{equation}
with the superscript $^\text{el}$ denoting the quantities for the electroquasistatic case. 

\subsection{Thermal Circuit and Equivalences}

In a thermal circuit, heat capacitances connect the circuit nodes to a common node (thermal ground) with zero temperature. This reflects the different nature of the term $\mathbf{M}_{\rho c}\dot{\mathbf{T}}$ in \eqref{eq:thermal_FIT} compared to the term $\widetilde{\mathbf{S}}\mathbf{M}_{\varepsilon}\widetilde{\mathbf{S}}^{\top}\dot{\mathbf{\Phi}}$ in \eqref{eq:EQS_FIT}.

With node $n+1$ as the thermal ground node, \eqref{eq:MNA_result} becomes
\begin{equation}
	\widetilde{\mathbf{A}}_\text{C}\mathbf{C}\widetilde{\mathbf{A}}_\text{C}^{\top}\dot{\widetilde{\mathbf{v}}} + \widetilde{\mathbf{A}}_\text{R}\mathbf{G}\widetilde{\mathbf{A}}_\text{R}^{\top}\widetilde{\mathbf{v}} = -\widetilde{\mathbf{A}}_\text{I}\mathbf{I}_\text{s}(t)
	\label{eq:MNA_result_extended}
\end{equation}
with $\widetilde{\mathbf{A}}_\text{C}\in\{-1,0,1\}^{(n+1)\times b_\text{C}}$, $\widetilde{\mathbf{A}}_\text{R}\in\{-1,0,1\}^{(n+1)\times b_\text{R}}$, $\widetilde{\mathbf{A}}_\text{I}\in\{-1,0,1\}^{(n+1)\times b_\text{I}}$ and $\widetilde{\mathbf{v}}\in\mathbbm{R}^{n+1}$ being the expansion of the corresponding quantities by this additional node. To mirror the structure of the discretized heat equation \eqref{eq:thermal_FIT}, we set the potential $v_\text{gnd}$ of the thermal ground node to zero and choose
\begin{align*}
	\begin{aligned}
		\widetilde{\mathbf{A}}_\text{C}^{\top} &= [\mathbbm{I} \quad -\mathbbm{1}],\\
		\widetilde{\mathbf{A}}_\text{I}^{\top} &= [\mathbbm{I} \quad -\mathbbm{1}]
	\end{aligned}
	&
	\begin{aligned}
		\\
		\quad \text{and} 
	\end{aligned}
	&
	\hspace{-30pt}
	\begin{aligned}
		\quad \widetilde{\mathbf{A}}_\text{R}^{\top} &= [\mathbf{A}_\text{R} \quad \mathbf{0}],\\
		\widetilde{\mathbf{v}}^{\top} &= [\mathbf{v}\quad v_\text{gnd}].
	\end{aligned}
\end{align*}
Here, $\mathbbm{I}$ is the $n\times n$ identity matrix and $\mathbbm{1}$ is a vector of ones with dimension $n$. If we then omit line $n+1$ in~\eqref{eq:MNA_result_extended}~(as we would do with a ground node in an electric circuit), we end up with \eqref{eq:MNA_result} again and obtain
\begin{equation}
	\begin{aligned}
		\quad \mathbf{A}_\text{C}^\text{th} &= \mathbf{A}_\text{I}^\text{th} = \mathbbm{I}, \,\,\, \mathbf{A}_\text{R}^\text{th} = \widetilde{\mathbf{S}}, \,\,\, \mathbf{C}^\text{th} = \mathbf{M}_{\rho c},\\
		\mathbf{G}^\text{th} &= \mathbf{M}_{\lambda}, \,\,\,  \mathbf{I}_\text{s}^\text{th} = -\mathbf{Q}_\text{el} \,\,\, \text{and} \,\,\, \mathbf{\mathbf{v}}^\text{th} = \mathbf{T},
		\label{eq:thermal_equivalence}
	\end{aligned}
\end{equation}
with the superscript $^\text{th}$ denoting the quantities for the thermal problem. 

The equivalences between the FIT formulation~\eqref{eq:EQS_FIT}~and~\eqref{eq:thermal_FIT} and the circuit formulation \eqref{eq:MNA_result} as shown above are the main motivation for this paper. The relations \eqref{eq:EQS_equivalence} and \eqref{eq:thermal_equivalence} allow to derive an equivalent SPICE netlist from any given 3D problem discretized by the FIT. The detailed methodology and numerical examples are discussed in the remaining sections of the paper. \section{SPICE Netlist Generation}
\label{sec:methodology}

This section deals with the implementation details for the automated netlist generation. First, the topology of the generated network is described for the EQS problem including the extraction of the Joule loss term. Then, the topology of the thermal network with inclusion of the Joule loss term is discussed. In the second part of the section, the approach for nonlinear material relations is shown.

\subsection{Circuit Topology}

\begin{figure}[b]
	\begin{subfigure}[b]{.5\columnwidth}
		\centering
		\includegraphics{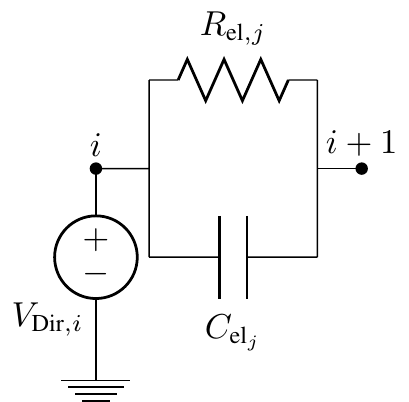}
 		\caption{Electrical circuit}
		\label{fig:electrical_circuit}
	\end{subfigure}
	\hspace{-0.5cm}
	\begin{subfigure}[b]{.49\columnwidth}
		\centering
		\includegraphics{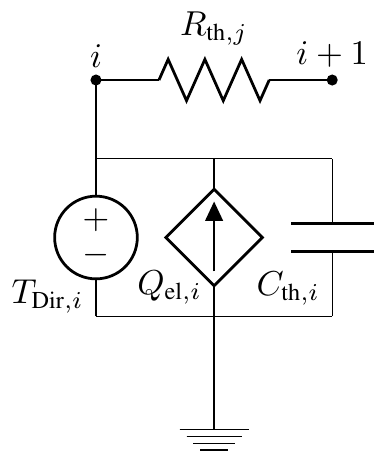}
 		\caption{Thermal circuit}
		\label{fig:thermal_circuit}
	\end{subfigure}
	\caption{Equivalent circuit model for example edge $L_j$ between nodes $i$ and $i+1$.}
\end{figure}

For the EQS problem, the degrees of freedom in the FIT are the potentials $\mathbf{\Phi}$ allocated at the primary nodes. In the MNA, these potentials are placed on the nodes of the network and thus remain nodal values. From the previous section, we know that the branch voltages of capacitive and resistive branches are equal, cf. \eqref{eq:voltage_definition} and \eqref{eq:EQS_equivalence}. Therefore, these elements are placed in parallel.  With the definition of the voltages $\mathbf{V}_\text{R}^\text{el} = \mathbf{V}_\text{C}^\text{el} = \widetilde{\mathbf{S}}^{\top}\mathbf{\Phi}$ as the difference of two nodal potentials, the resistive and capacitive elements are located on the branches between the nodes. From the pair-wise equality of $\mathbf{M}_{\sigma}$ and $\G$ as well as $\mathbf{M}_{\varepsilon}$ and $\mathbf{C}$, the entries $\mathbf{M}_{\sigma;j,j}$ and $\mathbf{M}_{\varepsilon;j,j}$ from the material matrices are the values for the lumped conductances and capacitances on circuit branch $L_j$, respectively. To impose inhomogeneous Dirichlet boundaries, voltage sources between a Dirichlet node and ground have to be defined. \figref{fig:electrical_circuit} shows the structure of the described electrical circuit for one example branch between nodes $i$ and $i+1$.

Since each branch contains resistive elements, the thermal losses need to be calculated on every branch. For branch $L_j$, this is achieved by the evaluation of the branch voltage $U_j$ and the current $I_j$ giving
\begin{equation*}
	\widehat{Q}_{\text{el},j} = U_j I_j.
\end{equation*}
Note that this thermal power $\widehat{Q}_{\text{el},j}$ coincides with the one calculated in the field model discussed in section \ref{sec:field_problem}. Writing the different $\widehat{Q}_{\text{el},j}$ in a vector, it becomes $\widehat{\mathbf{Q}}_{\text{el}}$ and leads to the thermal power $\mathbf{Q}_\text{el}$ on the dual volumes as given by \eqref{eq:Qel_definition}.

For the thermal problem, the structure of the conductive part ($\widetilde{\mathbf{S}}\mathbf{M}_{\lambda}\widetilde{\mathbf{S}}^{\top} \mathbf{T}$) is equal to the structure of the electrical problem. Therefore, as done for the electrical problem, conductive elements are placed on the branches between two circuit nodes. On the other hand, the structure of the capacitive part ($\mathbf{M}_{\rho c} \dot{\mathbf{T}}$) differs slightly from the EQS structure. Since the identity matrix is chosen for the incidence matrix of the capacitive branches ($\mathbf{A}_\text{C}^\text{th} = \mathbbm{I}$), the voltage $\mathbf{V}_\text{C} = \mathbbm{I}\mathbf{T}$ is not defined by temperature differences but by the absolute values of the temperatures (with respect to a reference temperature). If the temperature values are now allocated at the nodes of the thermal network, a capacitive connection from every thermal network node to the thermal ground as introduced in section \ref{sec:circuit_theory} is obtained. The values for these capacitances correspond to the values on the diagonal of $\mathbf{M}_{\rho c}$, cf. \eqref{eq:thermal_equivalence}. To include inhomogeneous Dirichlet conditions, temperature sources are modeled by voltage sources between the Dirichlet nodes and ground as done for the electrical circuit. Furthermore, the mapping of the heat powers $\mathbf{Q}_\text{el}$ to a circuit definition is given by its incidence matrix $\mathbf{A}_\text{I}^\text{th} = \mathbbm{I}$. Therefore, the corresponding current sources are placed between each circuit node and thermal ground. In \figref{fig:thermal_circuit}, these observations are summarized for a thermal example edge.

\subsection{Nonlinear Materials}
\label{sec:nonlinear}

In practice, all material properties may depend on the temperature. In this section, the dominating nonlinearity is assumed to be the electric conductivity $\sigma$. However, other dependencies can be incorporated analogously. If an inhomogeneous material distribution is given in the domain of interest, the average conductivity has to be found as presented in section \ref{sec:field_problem}. To correctly consider the nonlinearity in SPICE, this averaging needs to be carried out at runtime in the circuit solver. Therefore, a netlist with nonlinear entries is the result. 

If we consider the average temperature of an edge $\overline{T}_j$, the conductivity of the surrounding primary cells $V_p$ needs to be evaluated according to its nonlinear definition. In the simplest case, the temperature dependence of an isotropic conductivity is expressed via the resistivity, i.e.,
\begin{equation*}
	\rho_p(\overline{T}_j) = \frac{1}{\sigma_p(\overline{T}_j)} = \rho_{0,p}\left(1+\alpha_{p}\left(\overline{T}_j-T_0\right)\right).
\end{equation*}
In this equation, $\rho_{0,p}$ is the resistivity at the reference temperature $T_0$ and $\alpha_p\in\mathbbm{R}$ the temperature coefficient depending on the material of cell $V_p$. Once the nonlinear functions have been evaluated, the average conductivity $\overline{\sigma}_j$ of edge $L_j$ can be found as given by \eqref{eq:average_sigma}. Then, the nonlinear conductance $R_j(\overline{T}_j)$ is established by the length $\left|L_j\right|$ of primary edge $L_j$ and the area $|\widetilde{A}_j|$ of dual facet $\widetilde{A}_j$, giving
\begin{equation*}
	R_j(\overline{T}_j) = \frac{1}{\overline{\sigma}_j(\overline{T}_j)} \frac{|\widetilde{A}_j|}{\left|L_j\right|}.
\end{equation*}
Note that more complex material laws, non-equidistant grids and a more sophisticated averaging can be implemented analogously.

A pseudocode for the generation of electrothermal netlists with nonlinearities is shown in Algorithm~\ref{alg:netlist_pseudocode}.
\begin{algorithm}
	\caption{SPICE netlist generation algorithm}
	\label{alg:netlist_pseudocode}
	\begin{algorithmic}[1]
		\For{edge $L_j$ between primary nodes $i$ and $k$}
		\State write R\_el($j$) node($i$) node($k$) $R_j(\overline{T}_j)$
			\State write C\_el($j$) node($i$) node($k$) $\mathbf{M}_{\varepsilon}(j,j)$
			\State write R\_th($j$) node($i$) node($k$) $\mathbf{M}_{\lambda}^{-1}(j,j)$
			\State write C\_th($i$) gnd node($i$) $\mathbf{M}_{\rho c}(i,i)$
			\State write I\_Loss gnd node($i$) $Q_{\text{el},i}$
			\If{$i$ is electric Dirichlet node}
				\State write Vdir\_el($i$) node($i$) gnd $V_{\text{Dir},i}$
			\EndIf
			\If{$i$ is thermal Dirichlet node}
				\State write Vdir\_th($i$) node($i$) gnd $T_{\text{Dir},i}$
			\EndIf
		\EndFor
	\end{algorithmic}
\end{algorithm} \section{Numerical Validation}
\label{sec:numerical_example}

\subsection{Benchmark Example}

\begin{figure}[t]
	\vspace{-0.5em}
	\centering
	\includegraphics{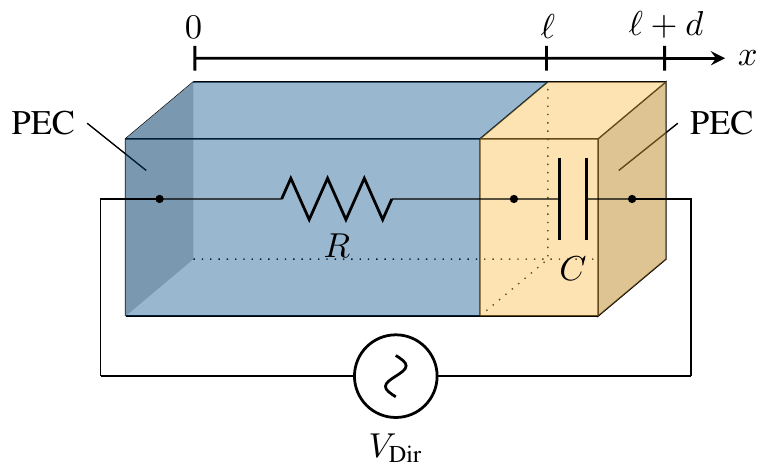}
 	\caption{Geometry of the benchmark example. A resistive part of length $\ell$ and a capacitive part of length $d$ is modeled and excited with a sinusoidal voltage source imposed as Dirichlet conditions.}
	\label{fig:benchmark_structure}
\end{figure}

For validation of the presented methodology, a benchmark cuboid of dimensions ${\SI{4}{mm}\times\SI{1}{mm}\times\SI{1}{mm}}$ is selected. It is composed of two different regions as shown in \figref{fig:benchmark_structure}. The left part of length $\ell=\SI{3}{mm}$ is chosen to be mainly resistive while the right part of length $d=\SI{1}{mm}$ is mainly capacitive. The material properties of the two regions are summarized in Table \ref{tab:material_properties}. On the boundaries in $x$-direction, Perfect Electric Conducting (PEC) facets are assumed. At the left boundary, a sinusoidal electric Dirichlet condition with \SI{1}{kV} amplitude and a frequency of \SI{76.9}{kHz} is imposed. At the right boundary, a potential of \SI{0}{V} is applied. Homogeneous Neumann conditions are chosen for the remaining boundaries. Thermally, all boundaries are set to homogeneous Neumann (thus adiabatic) conditions. For the initial conditions, all electric and thermal non-Dirichlet nodes are set to zero.

\begin{table}[t]
 \centering
 \caption{Material properties of the benchmark example.}
 \label{tab:material_properties}
 \begin{tabular}{|c|c|c|} \hline
	Property & $0<x<\ell$ & $\ell<x<\ell+d$ \\ \hline
	$\sigma$ $\left(\si{S\per m}\right)$ & \num{3} & \num{0}\\
	$\varepsilon_\text{r}$ & \num{1} & \num{1.13e5} \\
	$\lambda$ $\left(\si{W/K/m}\right)$ & \num{400} & \num{400} \\
	$\rho c$ $\left(\si{J\per K\per m^3}\right)$ & 8000 & 8000 \\ \hline
 \end{tabular}
 \vspace{-0.5em}
\end{table}

From a circuit point of view, the electric Dirichlet condition is modeled by a voltage source $V_\text{Dir}$ as the potential difference between the facets. Furthermore, the resistive material is modeled by a single resistance and the capacitive material by a single capacitor. However, the thermal properties of the two materials are assumed to be equal.

\begin{figure}[b]
	\centering
	\includegraphics{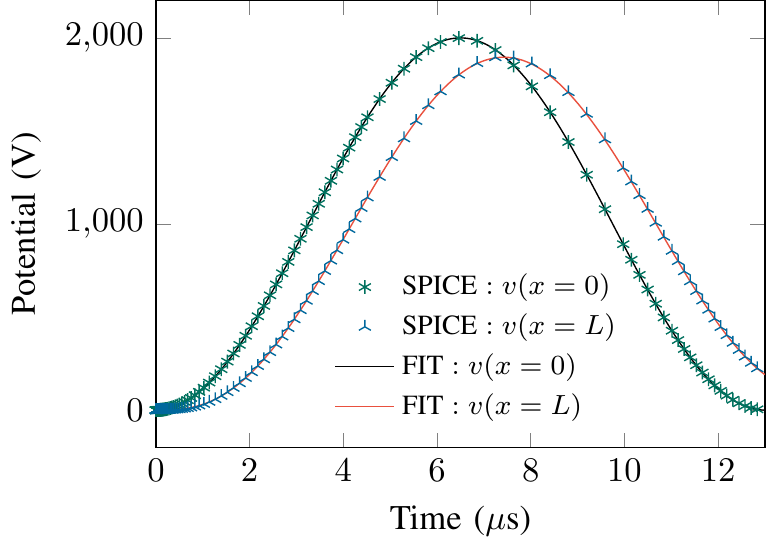}
 	\caption{Electric potential at selected points of the benchmark example calculated by the SPICE (crosses) and FIT (solid line) simulation.} 
	\label{fig:benchmark_potential}
\end{figure}

\begin{figure}[t]
	\centering
	\includegraphics{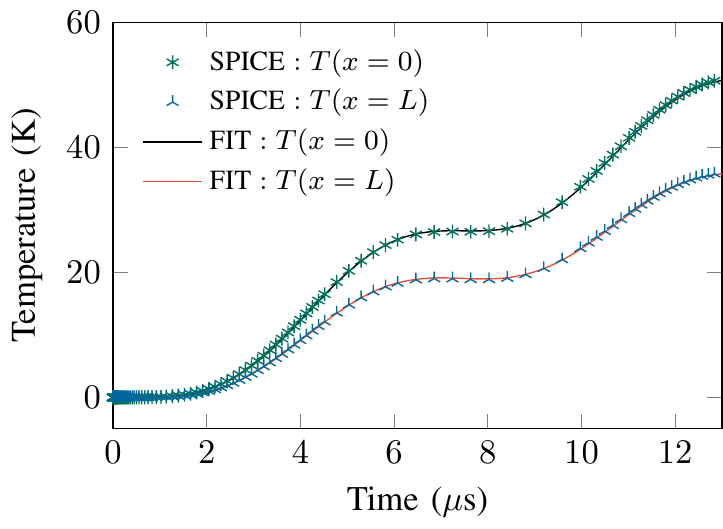}
 	\caption{Temperature at selected points of the benchmark example as the result of the SPICE (crosses) and FIT (solid line) simulation.}
	\label{fig:benchmark_temperature}
\end{figure}

After setting up the model with its material properties and boundary conditions, Algorithm \ref{alg:netlist_pseudocode} generates an equivalent electrothermal netlist. Then, using this netlist, a SPICE simulation is carried out and compared with the electrothermal FIT simulation. The electrical simulation result is shown in \figref{fig:benchmark_potential}. From the obtained potentials $\mathbf{\Phi}$, the thermal loss term $\mathbf{Q}_\text{el}$ is calculated. These losses are then coupled into the heat equation yielding the resulting temperatures as shown in \figref{fig:benchmark_temperature}.

From all three figures, it is clearly seen that a very good agreement of the SPICE and FIT simulation is achieved. The relative error norm for the obtained temperature is calculated as
\begin{equation}
	\text{error} = \frac{\max_i||\mathbf{T}_\text{SPICE}(t_i)-\mathbf{T}_\text{FIT}(t_i)||_2}{\max_i||\mathbf{T}_\text{FIT}(t_i)||_2} \approx \SI{0.52}{\%},
	\label{eq:error}
\end{equation}
where $\mathbf{T}_\text{SPICE}$ and $\mathbf{T}_\text{FIT}$ are $n_t$ vectors of dimension $n$ with $n_t$ being the number of time steps. Since the spatial discretization is chosen identically and the generated netlist is a representation of the 3D field model without any further simplifications, the only error is resulting from the different time integrators.

\subsection{Chip Package}
\begin{figure}[b]
	\centering
	\includegraphics[width=0.6\columnwidth]{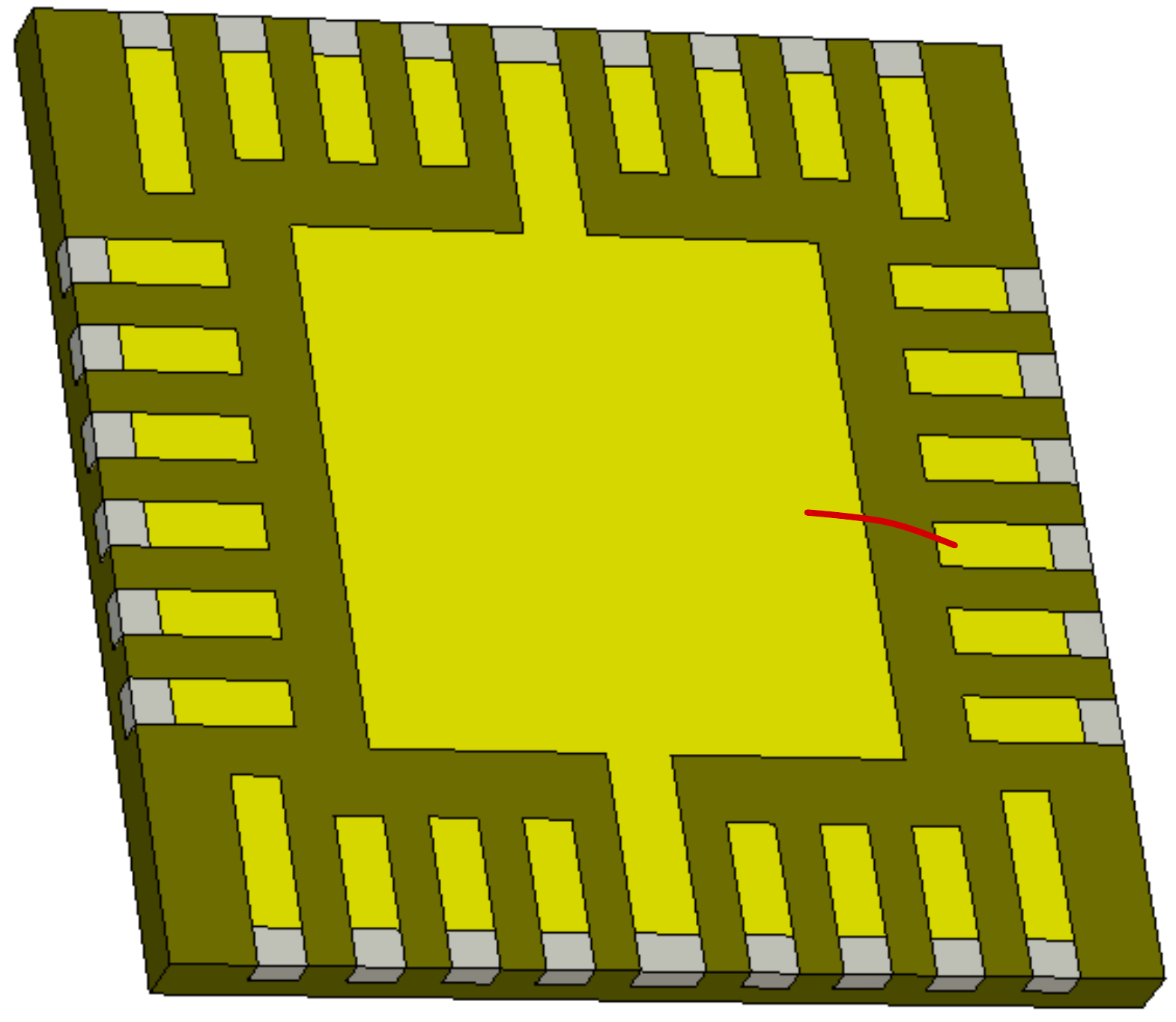}
	\caption{Chip package with one bonding wire.}
	\label{fig:chip}
\end{figure}

\begin{figure}[t]
	\centering
	\includegraphics{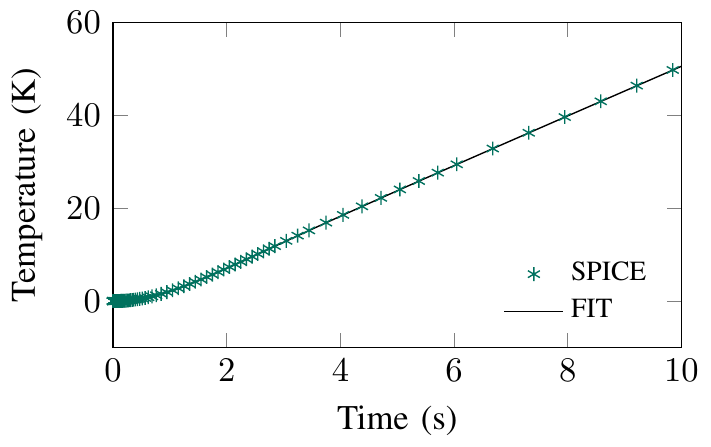}
 	\caption{Temperature at the hottest point of the microelectronic chip package obtained by SPICE simulation.}
	\label{fig:benchmarkChip}
\end{figure}

To apply the method to a more complex example, a microelectronic chip package as shown in \figref{fig:chip} is considered. Here, the same setting as in \cite{Casper_2016aa} is chosen. However, only one bonding wire connects one of the contacts to the chip. The bonding wire's electrical conductance is $G_{\text{bw}}^{\text{el}} = \SI{1}{S}$ and its thermal conductance $G_{\text{bw}}^{\text{th}} = \SI{1}{kW\per K}$. Furthermore, the relative permittivity for all materials are set to one while the thermal boundary conditions are adiabatic. To excite the system, an exponential voltage $V_0(t) = \SI{10}{V}\left[1-\exp(-t)\right]$ is applied over the bonding wire. Algorithm \ref{alg:netlist_pseudocode} is executed to generate the equivalent netlist. This is simulated using SPICE and the temperature of the hottest node of the domain is compared to the result obtained by FIT as shown in \figref{fig:benchmarkChip}. The error as calculated by \eqref{eq:error} is \SI{0.07}{\%}. Due to the mainly resistive character of this example, the resulting constant current leads to a constant heating of the considered node. 

 \section{Conclusions}
\label{sec:conclusion}

The automatic generation of netlists directly from an electrothermal 3D field model has been presented. Starting from a Finite Integration Technique~(FIT) discretization, an equivalent circuit description in terms of the Modified Nodal Analysis~(MNA) was shown for the electrical as well as for the thermal case. As it is a direct representation of the field model, no further simplifications apply when generating the netlist. For validation, the field solution of an electrothermal benchmark example and a complex 3D chip package problem have been compared to the SPICE solution based on the generated netlist. 

Next steps are the treatment of non-hexahedral meshes, arbitrary nonlinear materials and model order reduction of the resulting network to reduce the complexity of the obtained circuit and thus enable a more efficient circuit simulation.

\section{Acknowledgement}
The authors thank Abdul Moiz for his passionate work during the first implementation of the algorithm for the automatic netlist generation.	The work is supported by the European Union within FP7-ICT-2013 in the context of the \emph{Nano-electronic COupled Problems Solutions} (nanoCOPS) project (grant no. 619166), by the \emph{Excellence Initiative} of the German Federal and State Governments and the Graduate School of CE at TU Darmstadt.

\phantom{i}

\end{document}